\documentclass[%
 reprint,
superscriptaddress,
 amsmath,amssymb,
 aps,
]{revtex4-2}

\usepackage[utf8]{inputenc}
\usepackage{graphicx}
\usepackage{color}
\usepackage{physics}
\usepackage{braket}
\usepackage{comment}
\usepackage{mathrsfs}
\usepackage{enumerate}
\usepackage[version=4]{mhchem}
\usepackage[breaklinks=true]{hyperref}
\newcommand{\mr}[1]{\mathrm{#1}}
\newcommand{\ms}[1]{\mathscr{#1}}
\newcommand{\mcl}[1]{\mathcal{#1}}
\newcommand{\bbC}{\mathbb{C}}
\newcommand{\bbR}{\mathbb{R}}
\newcommand{\bbZ}{\mathbb{Z}}
\newcommand{\bbN}{\mathbb{N}}
\newcommand{\eff}{\mathrm{eff}}

\newcommand{\imax}{\mathrm{max}}
\newcommand{\imin}{\mathrm{min}}
\newcommand{\poly}[1]{\mathrm{poly} \left( #1 \right)}
\date{\today}
\usepackage{amsthm}
\theoremstyle{definition}
\newtheorem{theorem}{Theorem}[]

\newtheorem*{theorem*}{Theorem}
\newtheorem*{proposition2*}{Proposition}

\begin{document}
\title{Optimal/Nearly-optimal simulation of multi-periodic time-dependent Hamiltonians}

\author{Kaoru Mizuta}
\email{kaoru.mizuta@riken.jp}
\affiliation{RIKEN Center for Quantum Computing (RQC), Hirosawa 2-1, Wako, Saitama 351-0198, Japan}

\begin{abstract}
Simulating Hamiltonian dynamics is one of the most fundamental and significant tasks for characterising quantum materials.
Recently, a series of quantum algorithms employing block-encoding of Hamiltonians have succeeded in providing efficient simulation of time-evolution operators on quantum computers.
While time-independent Hamiltonians can be simulated by the quantum eigenvalue transformation (QET) or quantum singularvalue transformation with the optimal query complexity in time $t$ and desirable accuracy $\varepsilon$, generic time-dependent Hamiltonians face at larger query complexity and more complicated oracles due to the difficulty of handling time-dependency.
In this paper, we establish a QET-based approach for simulating time-dependent Hamiltonians with multiple time-periodicity.
Such time-dependent Hamiltonians involve a variety of nonequilibrium systems such as time-periodic systems (Floquet systems) and time-quasiperiodic systems.
Overcoming the difficulty of time-dependency, our protocol can simulate the dynamics under multi-periodic time-dependent Hamiltonians with optimal/nearly-optimal query complexity both in time $t$ and desirable accuracy $\varepsilon$, and simple oracles as well as the optimal algorithm for time-independent cases.
\end{abstract}

\maketitle

\section{Introduction}\label{Sec:Introduction}

Hamiltonian simulation, that is, constructing time-evolution operators by a set of elementary quantum gates, is one of the most important tasks of quantum computers \cite{feynman1982}.
For instance, it can be exploited for reproducing the solution of Schr\"{o}dinger equation \cite{Lanyon2011-qo,Smith2019-zz,Arute2020-qr} or identifying energy eigenvalues and eigenstates by quantum phase estimation algorithms \cite{Yu_Kitaev1995-oi,Cleve1998-ou,Du2010-hl,Lanyon2010-wk,OMalley2016-dh}, both of which are classically hard but significant problems in condensed matter physics and quantum chemistry.
Therefore, accurate and efficient Hamiltonian simulation has been the central issue in quantum computation, addressed by various strategies, such as Trotterization \cite{Lloyd1996-ko,Abrams1997-jb,Sornborger1999-oq,Childs2021-trotter}, and variational quantum compling \cite{Cirstoiu2020-kk,Commeau2020-tf,Mizuta2022-LVQC}.

Recently, the so-called qubitization technique \cite{Low2019-qubitization} has become one of the most promising protocols for simulating large-scale Hamiltonians, which can achieve high accuracy with much smaller cost than the standard way Trotterization.
Based on quantum eigenvalue transformation (QET) or quantum singular value transformation (QSVT) \cite{Gilyen2019-qsvt,MartynPRXQ-grand-unif}, it organizes the time-evolution $U(t)=e^{-iHt}$ under time-independent Hamiltonians $H$ with some ancillary qubits and queries to a block-encoding.
Significantly, it works with the theoretically-best query complexity both in time $t$ and acceptable error $\varepsilon$ \cite{Berry2015-gt,Low2017-QSP}.
However, when it comes to time-dependent Hamiltonians $H(t)$, we suffer from the difficulty of time-dependency as the time-evolution becomes
\begin{equation}
    U(t) = \mcl{T} \exp \left( -i \int_0^t \dd t^\prime H(t^\prime) \right).
\end{equation}
We cannot generally use QET or QSVT which are only valid for one-variable functions of $H$, and implementing the time-ordered integration is not straightforward.
Although the truncated-Dyson-series algorithm \cite{Low2018-dyson,Kieferova2019-dyson} and others \cite{Berry2020-time-dep,Haah2021-time-dep,Chen2021-time-dep,Watkins2022-time-dep} deal with this task by discretizing the time, we need complicated oracles, larger query complexity, and larger ancilla systems compared to time-independent cases.
As far as we know, QET-based approaches for simulating time-dependent systems as efficiently as time-independent systems have been limited to time-periodic Hamiltonians such that $H(t+T)=H(t)$ with some period $T$ \cite{Mizuta2022optimal}.
It is therefore important to explore what kind of time-dependent Hamiltonians can be simulated as efficiently as time-independent systems by explicitly constructing algorithms.

In this paper, we formulate an efficient QET-based approach for simulating multi-periodic time-dependent Hamiltonians in the form of
\begin{equation}
    H(t) = \sum_{\vec{m}} H_{\vec{m}} e^{-i \vec{m} \cdot \vec{\omega} t}, \quad \vec{m} \in \bbZ^n, \quad \vec{\omega} \in \bbR^n.
\end{equation}
Not only they include time-periodic Hamiltonians by $n=1$, they provide a significant class of nonequilibrium systems known as time-quasiperiodic systems \cite{Chu2004-we-multi}.
In a similar way to the optimal algorithm for time-periodic Hamiltonians \cite{Mizuta2022optimal}, we establish a protocol to simulate the time-evolution without relying on neither time-ordered product nor Dyson-series expansion.
Instead, we prepare an ancillary quantum state $\ket{\vec{l}}$ which labels multiple Fourier indices $\vec{l} \in \bbZ^n$ in the frequency domain.
This procedure enables much simpler realization of the time-evolution by the qubitization technique for a certain time-independent Hamiltonian, called the effective Hamiltonian.
As a result, with employing the block-encoding of each Fourier component $H_{\vec{m}}$, we achieve the query complexity as the computational cost in the form of
\begin{equation}
    \alpha t + \omega t \times o \left( \log (\omega t /\varepsilon) \right).
\end{equation}
Here, we have $\alpha \in \poly{N}$ and $\omega \in \order{N^0}$ for typical $N$-site quantum many-body systems.
This scaling is optimal in time $t$ and nearly-optimal in allowable error $\varepsilon$. 
Importantly, its additive form implies that multi-periodic Hamiltonians can be simulated almost as efficiently as time-independent systems, which reaches the theoretically-best scaling \cite{Low2019-qubitization}.

Our algorithm for multi-periodic Hamiltonians extends a class of time-dependent systems that can be efficiently addressed based on QET, in spite of the difficulty of handling time-dependency on quantum circuits.
Furthermore, multi-periodic time-dependent Hamiltonians themselves have been attractive platforms for various nonequilibrium phenomena \cite{Chu2004-we-multi,Martin2017-multi,Crowley2019-multi,Zhao2019-multi,Else2020-multi}, such as frequency conversion in multiple-light-irradiated materials \cite{Martin2017-multi}, and topological or time-quasicrystalline phenomena in time-quasiperiodic systems \cite{Else2020-multi}.
Therefore, our result will enhance potential of quantum computers toward condensed matter physics and quantum chemistry.

The rest of this paper is organized as follows.
In Section \ref{Sec:Preliminaries}, we briefly review qubitization for simulating time-independent Hamiltonians.
After summarizing our main result in Section \ref{Sec:Summary_paper}, we provide its detailed derivation from Section \ref{Sec:Floquet_Hilbert} to Section \ref{Sec:Algorithm}.
We conclude this paper in Section \ref{Sec:Discussion}.

\section{Brief review on qubitization}\label{Sec:Preliminaries}

In this section, we briefly review Hamiltonian simulation of time-independent systems based on QET, which is so-called the qubitization technique \cite{Low2019-qubitization}.

We first construct block-encoding which embeds time-independent Hamiltonians $H$ into a unitary gate $O$ such that
\begin{equation}\label{Eq2:block_encode_qubitization}
    \braket{0|O|0}_a = \frac{H}{\alpha}, \quad \alpha > 0.
\end{equation}
The quantum state $\ket{0}_a$ denotes a trivial reference state of an $n_a$-qubit ancilla system.
Due to the unitarity of $O$ implying $\norm{O}=1$ ($\norm{\cdot}$; operator norm), the denominator $\alpha$ should satisfy $\alpha \geq \norm{H}$.
As a result, $\alpha \in \poly{N}$ is satisfied for typical $N$-site quantum many-body systems.
For instance, let us consider a Hamiltonian given by a linear combination of unitaries (LCU),
\begin{equation}
    H = \sum_{j=1}^{j_\imax} \alpha_j U_j, \quad \alpha_j \geq 0, \quad \text{$U_j$; unitary},
\end{equation}
where the number of terms $j_\imax$ typically satisfies $j_\imax \in \poly{N}$.
We can organize its block-encoding by
\begin{eqnarray}
    && O = (G_a \otimes I)^\dagger \left( \sum_{j=1}^{j_\imax} \ket{j}\bra{j}_a \otimes U_j \right) (G_a \otimes I), \label{Eq2:oracle_LCU} \\
    && G_a \ket{0}_a  = \sum_{j=1}^{j_\imax} \sqrt{\frac{\alpha_j}{\alpha}} \ket{j}_a, \quad \alpha = \sum_{j=1}^{j_\imax} \alpha_j, \label{Eq2:prepare_LCU}
\end{eqnarray}
with the number of ancilla qubits $n_a = \lceil \log_2 j_\imax \rceil \in \order{\log N}$.
LCUs cover various spin systems and fermionic systems in condensed matter physics and quantum chemistry \cite{Babbush2018-hz,Low2019-qubitization,Gilyen2019-qsvt}.

The next step of the qubitization is to execute QET with the oracle $O$ \cite{Gilyen2019-qsvt}.
We use the phase rotation on the ancilla system given by
\begin{equation}
    R(\phi) = e^{i\phi (2\ket{0}\bra{0}_a - I_a)} \otimes I, \quad \phi \in [0,2\pi),
\end{equation}
which yields $\order{n_a}$ elementary gates.
Then, we organize a unitary operation $W_Q$ with $\order{Q}$-times usage of $R(\phi)$ and $O$ (or $O^\dagger$), and also with $\order{1}$ additional qubits.
By properly tuning the parameter set $\phi_1, \hdots, \phi_{\order{Q}}$ in the rotations $R(\phi)$ using the technique of quantum signal processing \cite{Low2017-QSP}, the unitary gate $W_Q$ enables to execute broad classes of degree-$Q$ polynomial functions of $H$ as
\begin{equation}
    \braket{0|W_Q|0}_{a^\prime} = f_Q(H) = \sum_{k=0}^Q c_k H^k, \quad c_k \in \bbC,
\end{equation}
with an $\{ n_a + \order{1} \}$-qubit ancilla system $a^\prime$.
For Hamiltonian simulation, we employ a degree-$Q$ polynomial $f_Q(H)$ approximately giving $e^{-iHt}$ (e.g., truncated Jacobi-Anger expansion or Taylor-series expansion).
As a result, we can apply $U(t)=e^{-iHt}$ to arbitrary quantum states $\ket{\psi}\in \mcl{H}$ by the unitary gate $W_Q$ as
\begin{equation}
    W_Q \ket{0}_{a^\prime} \ket{\psi} = \ket{0}_{a^\prime} e^{-iHt} \ket{\psi} + \order{\varepsilon_Q}, \quad \varepsilon_Q = \left( \frac{\alpha t}{Q}\right)^Q,
\end{equation}
where the error $\varepsilon_Q$ arises from degree-$Q$ polynomial approximation of $e^{-iHt}$.

The cost of Hamiltonian simulation is evaluated by the resource for $W_Q$ with achieving a desirable error $\varepsilon$ as
\begin{equation}
    \varepsilon_Q = \left( \frac{\alpha t}{Q}\right)^Q \leq \varepsilon.
\end{equation}
The number $Q$ required for satisfying this inequality is obtained by the notion of the Lambert W function $W(x)$ \cite{Hoorfar2008-lambert}, whose consequence dictates
\begin{equation}\label{Eq2:ineq_Lambert_W}
    \left(\frac{\kappa}{x}\right)^x \leq \eta, \quad ^\forall x \geq e \kappa + \frac{4 \log (1/\eta)}{\log (e+\kappa^{-1}\log(1/\eta))},
\end{equation}
for any $\kappa \in [0,\infty)$ and $\eta \in (0,1]$ \cite{Gilyen2019-qsvt}.
Thus, it is sufficient to choose $Q$, which gives the number of the oracle $O$ and the rotation $R(\phi)$ used for $W_Q$, by
\begin{equation}\label{Eq2:query_qubitization}
    Q \in \order{ \alpha t + \frac{\log (1/\varepsilon)}{\log (e+(\alpha t)^{-1} \log (1/\varepsilon))} },
\end{equation}
to implement $U(t)$ with the desirable error $\varepsilon$.
Qubitization requires following resources for simulating time-independent Hamiltonians $H$;
\begin{itemize}
    \item Number of ancilla qubits $n_{a^\prime}$; $ n_a+\order{1}$
    \item Query complexity; $Q$ [Eq. (\ref{Eq2:query_qubitization})]
    \item Additional gate per query $\mcl{N}_a$; $\order{n_a}$
\end{itemize}
Query complexity means the number of the oracle $O$ in the algorithm, which dominantly determines the computational cost.
Additional gates per query comes from the resource for each phase rotation $R(\phi)$.
When the oracle $O$ can be executed by $\mcl{N}_o$ elementary gates, the algorithm needs $\order{Q \mcl{N}_o + Q \mcl{N}_a}$ gates.

Significantly, linear scaling in time as $\order{t}$ and logarithmic scaling in accuracy as $\order{\log(1/\varepsilon) / \log \log (1/\varepsilon)}$ are both known to be optimal for the query complexity of Hamiltonian simulation \cite{Berry2015-gt,Low2017-QSP}.
The query complexity of qubitization, given by Eq. (\ref{Eq2:query_qubitization}) in an additive way, achieves the best scaling in time and accuracy for time-independent systems.
In contrast, while QET allows efficient quantum algorithms by proper polynomial approximation, it is unsuited for time-dependent systems since their time-evolution $U(t)$ is not simply expressed by univariate polynomial functions. 
This results in larger resources for Hamiltonian simulation of time-dependent systems like the truncated-Dyson-series algorithms \cite{Low2018-dyson,Kieferova2019-dyson}.
Nevertheless, we will show that we can exploit this technique for time-dependent systems if we impose multiple time-periodicity.
This leads the optimal/nearly-optimal cost of our algorithm having an additive form, as discussed later.

\begin{table*}
    \centering
    \begin{tabular}{|c|c|c|c|}
         &  Ancilla qubits & Query complexity & Additional gates per query \\ \hline \hline
       \begin{tabular}{c} Time-independent $H$ \\ (Qubitization \cite{Low2019-qubitization}) \end{tabular} & $n_a+\order{1}$ &
      $\alpha t + \frac{\log (1/\varepsilon)}{\log (e + (\alpha t)^{-1} \log (1/\varepsilon))}$
        & $\order{n_a}$
       \\ \hline
       \begin{tabular}{c} Multi-periodic $H(t)$ \\ ($\omega t \in \order{1}$, Thm. \ref{Thm:algorithm_Order1}) \end{tabular} & $n_a+\order{\log (\gamma t) + \log \log (1/\varepsilon)}$ & $\alpha t + o (\log (1/\varepsilon))$ & $\order{n_a+\log (\gamma t)+\log \log (1/\varepsilon)}$ \\ \hline
       \begin{tabular}{c} Multi-periodic $H(t)$ \\ ($\omega t \in \Omega (1)$, Thm. \ref{Thm:algorithm_Omega1}) \end{tabular} & $n_a+\order{\log (\gamma/\omega) + \log \log (\omega t/\varepsilon)} $ & $ \alpha t +  \omega t \times o(\log (\omega t /\varepsilon)) $ & $\order{n_a+\log (\gamma/\omega) + \log \log (\omega t/\varepsilon)}$
       \\ \hline
       \begin{tabular}{c} Time-dependent $H(t)$ \\ (Dyson series \cite{Low2018-dyson,Kieferova2019-dyson}) \end{tabular} & $n_a+\order{\log \{ (\gamma \omega t / \alpha + \alpha t) / \varepsilon \}}$ & $\alpha t \frac{\log(\alpha t/\varepsilon)}{\log \log (\alpha t /\varepsilon)}$ & $\order{n_a + \log \{ (\gamma \omega t / \alpha + \alpha t) / \varepsilon \}}$ \\ \hline
    \end{tabular}
    \caption{Comparison of computational resource for Hamiltonian simulation. Our results on multi-periodic Hamiltonians are based on Theorems \ref{Thm:algorithm_Order1} and \ref{Thm:algorithm_Omega1}. We note that we replace definitions of some symbols by those having similar scales to make the comparison easier. Precisely speaking, the symbol $\alpha$ denotes the energy scale of the whole Hamiltonian $\alpha \gtrsim \norm{H}$ or $\alpha \gtrsim \max_t (\norm{H(t)})$. The symbol $\gamma$ is used for the energy scale of time-dependent terms, i.e., $\gamma \omega \sim \max_t (\norm{\dv{t}H(t)})$.}
    \label{Table:comparison_algorithms}
\end{table*}

\section{Summary of Main results}\label{Sec:Summary_paper}

\subsection{Setup}\label{Subsec:Setup}

We first specify time-dependent systems of interest throughout the paper.
We consider a quantum system on a finite-dimensional Hilbert space $\mcl{H}$.
With the frequency of multiple drives by a $n$-dimensional vector $\vec{\omega} = (\omega_1,\omega_2,\hdots,\omega_n)$ ($\omega_i > 0$), a Hamiltonian on the Hilbert space $\mcl{H}$ is assumed to be written by
\begin{equation}\label{Eq3:Hamiltonian}
    H(t) = \sum_{\vec{m} \in M} H_{\vec{m}} e^{-i \vec{m} \cdot \vec{\omega} t},
\end{equation}
where $M$ is a finite set of $\bbZ^n$ and each $H_{\vec{m}}$ is a matrix on the Hilbert space $\mcl{H}$.
The hermiticity of $H(t)$ at every time imposes $H_{-\vec{m}} = H_{\vec{m}}^\dagger$.
We assume the satisfaction of
\begin{equation}
    n, |M| \in \order{1}, \quad \max_{\vec{m} \in M} (|\vec{m}|) \leq m_\imax \in \order{1}
\end{equation}
throughout the discussion.
We define the characteristic scale of frequency $\omega$ and the characteristic time scale $T$ by
\begin{equation}
    \omega = |\vec{\omega}|_1 = \sum_{i=1}^n \omega_i, \quad T = \frac{2\pi}{\omega},
\end{equation}
which we simply call frequency and period later.
We also impose that the frequency $\omega$ is an $\order{N^0}$ constant much smaller than the total energy scale of the Hamiltonian $\max_{t \in \bbR} \{ \norm{H(t)} \} \in \poly{N}$ for typical size-$N$ quantum many-body systems.

The above Hamiltonian can be re-described by a $n$-dimensional Fourier series expansion as $H(t) = \left. \overline{H}(\vec{x}) \right|_{\vec{x}=\vec{\omega}t}$ ($\vec{x} \in \bbR^n$), which is defined by
\begin{eqnarray}
    \overline{H}(\vec{x}) &=& \sum_{\vec{m} \in M} H_{\vec{m}} e^{-i \vec{\omega} \cdot \vec{x}}, \\
    H_{\vec{m}} &=& \int_{[0,2\pi)^n} \frac{\dd \vec{x}}{(2\pi)^n} \overline{H}(\vec{x}) e^{i \vec{m} \cdot \vec{x}}.
\end{eqnarray}
The Hamiltonian $\overline{H}(\vec{x})$ satisfies $\overline{H}(\vec{x}+2\pi \vec{e}_i)$ with a unit vector $\vec{e}_i$ in any direction $i=1,2,\hdots,n$.
In that sense, the above multi-periodic Hamiltonians $H(t)$ are natural extensions of time-periodic Hamiltonian, and said to have multiple time translation symmetry.
We also note that $H(t)$ becomes time-quasiperiodic if there exists a pair of $\omega_i$ and $\omega_j$ giving an irrational ratio $\omega_i/\omega_j$.
Although Eq. (\ref{Eq3:Hamiltonian}) is often used for describing time-quasiperiodic cases, we do not impose this condition.
Finally, we define $\gamma$ by
\begin{equation}\label{Eq3:def_gamma}
    \gamma = \sup_{\vec{x} \in [0,2\pi)^n} \left( \norm{\overline{H}(\vec{x})-H_{\vec{0}}}\right).
\end{equation}
This provides the upper bound on the total energy scale of time-dependent terms in multi-periodic Hamiltonians $H(t)$.

\subsection{Main results}\label{Subsec:Main_results}

We briefly show the main results of this paper here.
We organize a unitary circuit that applies a time-evolution operator $U(t)$ to an arbitrary initial state $\ket{\psi} \in \mcl{H}$.
First, we assume that the following oracles are given;
\begin{enumerate}
    \item A set of unitary gate $\{ O_{\vec{m}} \}_{\vec{m} \in M}$ giving a block-encoding of each Fourier component as
    \begin{equation}\label{Eq3:block_encode_Fourier}
        \bra{0} O_{\vec{m}} \ket{0}_a = \frac{H_{\vec{m}}}{\alpha_{\vec{m}}}, \quad \alpha_{\vec{m}} > 0,
    \end{equation}
    where $\ket{0}_a$ denotes a trivial reference state of an $n_a$-qubit ancilla system.
    
    \item A unitary gate $G_\mr{coef}$ giving an access to each coefficient $\alpha_{\vec{m}}$ as
    \begin{equation}\label{Eq3:def_G_coef}
        G_\mr{coef} \ket{\vec{0}}_b = \sum_{\vec{m} \in M} \sqrt{\frac{\alpha_{\vec{m}}}{\alpha}} \ket{\vec{m}}_b, \quad \alpha = \sum_{\vec{m} \in M} \alpha_{\vec{m}}.
    \end{equation}
    The system $b$ has $\order{\log (|M|)} = \order{1}$ qubits and can occupy quantum states spanned by $\{ \ket{\vec{m}} \}_{\vec{m} \in M}$.
    
    \item A unitary gate $G_\mr{freq}$ giving an access to each frequency $\omega_i$ as
    \begin{equation}\label{Eq3:def_G_freq}
        G_\mr{freq} \ket{0}_c = \sum_{i=0}^{n-1} \sqrt{\frac{\omega_{i+1}}{\omega}} \ket{i}_c,
    \end{equation}
    where the ancilla system $c$ possesses $\order{\log n} = \order{1}$ qubits.
\end{enumerate}
The block-encoding Eq. (\ref{Eq3:block_encode_Fourier}) requires $\alpha_{\vec{m}} \geq \norm{H_{\vec{m}}}$ for every $\vec{m}$, and we obtain
\begin{equation}\label{Eq3:ineq_alpha}
    \norm{H(t)} \leq \sum_{\vec{m} \in M} \norm{H_{\vec{m}}} \leq \alpha.
\end{equation}
The parameter $\alpha$ is therefore $\poly{N}$ for typical size-$N$ quantum many-body systems.
We also note that the oracle $O_{\vec{m}}$ usually yields more resource than the others $G_\mr{coef}$ and $G_\mr{freq}$ since they manipulate $\order{1}$ qubits and similar quantum gates are often included in $O_{\vec{m}}$ [e.g. See Eq. (\ref{Eq2:prepare_LCU}) for LCUs].
Comparing with the oracles used for qubitization given by Eq. (\ref{Eq2:block_encode_qubitization}), the difficulty of preparing oracles for our setups is essentially the same as that for time-independent systems.

Upon the above setup, our central result is construction of a unitary gate $W(t)$ which gives a time-evolution operator $U(t)$ under the multi-periodic Hamiltonian $H(t)$ to an arbitrary quantum state $\ket{\psi} \in \mcl{H}$ as
\begin{equation}
    W(t) \ket{0}_{a^\prime} \ket{\psi} = \ket{0}_{a^\prime} \{ U(t) \ket{\psi} \} + \order{\varepsilon}.
\end{equation}
Here, $\ket{0}_{a^\prime}$ is a trivial reference state of an ancilla system $a^\prime$.
The cost of Hamiltonian simulation is determined by the number of ancilla qubits for $\ket{0}_{a^\prime}$ and the query complexity, which means the number of the oracles $\{ O_{\vec{m}}, G_\mr{coef} \}$ employed for the unitary gate $W(t)$.
We summarize our results in Table \ref{Table:comparison_algorithms}, although the detailed discussion is left for Section \ref{Sec:Algorithm}.
We obtain efficient algorithms for two different time scales. 
The first one is $\order{1}$-period dynamics with $\omega t \in \order{1}$, while the second one is $\Omega(1)$-period dynamics in which we typically have $\omega t \gg 1$.
At first glance, we see that the number of ancilla qubits is just between the qubitization and the Dyson-series algorithm.
Significantly, the query complexity has optimal scaling both in time $t$ and inverse error $1/\varepsilon$, and takes an additive form of $\alpha t$ and $\log (1/\varepsilon)$.
This scaling is equal or close to the theoretically best one for time-independent systems.
As a matter of fact, the computational resource for multi-periodic Hamiltonians is essentially the same as that for time-periodic Hamiltonians \cite{Mizuta2022optimal}, where we set $n=1$ in Eq. (\ref{Eq3:Hamiltonian}).
Therefore, our result reveals that a larger class of time-dependent Hamiltonians including multi-periodicity can be simulated as efficiently as time-independent systems.

\section{Time-evolution from Floquet-Hilbert space}\label{Sec:Floquet_Hilbert}

In this section, we discuss how we compute the exact time-evolved state $U(t) \ket{\psi}$ for an arbitrary initial state $\ket{\psi}$, with keeping the error up to $\order{\varepsilon}$.
The principal strategy relies on a natural extension of the optimal algorithm for time-periodic Hamiltonians \cite{Mizuta2022optimal}.
Analogous to time-periodic systems, we introduce an ancilla quantum system $f$ labeling Fourier indices $\vec{l}$ by $\{ \ket{\vec{l}}_f \}_{\vec{l} \in [L]^n}$, where the domain $[L]^n$ denotes a subset of $\bbZ^n$ defined by
\begin{equation}
    [L]^n = \{ -L + 1 , -L+2, \hdots, L \}^n.
\end{equation}
We refer to a space $\mcl{H}_\mr{F}^L$ defined by
\begin{equation}
    \mcl{H}_\mr{F}^L = \mr{span} \left\{ \ket{\vec{l}}_f \right\}_{\vec{l} \in [L]^n} \otimes \mcl{H}
\end{equation}
as a Floquet-Hilbert space.
We also define an effective Hamiltonian $\ms{H}^L$ acting on the Floquet-Hilbert space by
\begin{eqnarray}
    \ms{H}^{L} &=& \sum_{\vec{m} \in M} \mr{Add}_{\vec{m}}^{L} \otimes H_{\vec{m}} - \ms{H}_\mr{LP}^L, \label{Eq4:def_H_eff} \\
    \ms{H}_\mr{LP}^L &=& \sum_{\vec{l} \in [L]^n} \vec{l} \cdot \vec{\omega} \ket{\vec{l}}\bra{\vec{l}}_f \otimes I. \label{Eq4:def_H_linear_potential}
\end{eqnarray}
Here, the unitary operator $\mr{Add}_{\vec{m}}^L$ executes addition of $\vec{l}$ as
\begin{equation}\label{Eq4:addition_torus}
    \mr{Add}_{\vec{m}}^L = \sum_{\vec{l} \in [L]^n} \ket{\vec{l} \oplus \vec{m}}\bra{\vec{l}}_f,
\end{equation}
where each $l_i \oplus m_i \in [L]$ is defined modulo $2L$.
The term $\ms{H}_\mr{LP}^L$ represents a linear potential for Fourier indices.
We note that the form of the effective Hamiltonian $\ms{H}^L$ is different from that of the standard many-mode Floquet theory \cite{Chu2004-we-multi}, in that $\ms{H}^L$ includes unphysical excitations $\ket{\vec{l} \oplus \vec{m}}\bra{\vec{l}} \otimes H_{\vec{m}}$ with $\vec{l} \oplus \vec{m} \neq \vec{l} + \vec{m}$.

To construct the optimal algorithm for multi-periodic Hamiltonians $H(t)$, we first establish another expression of the time-evolved state $U(t) \ket{\psi}$.
This expression does not rely on Dyson-series expansion, but instead exploits the Floquet-Hilbert space and the effective Hamiltonian $\ms{H}^L$, as we discuss in Section \ref{Subsec:time_evol_repr}.
Next, we construct a quantum algorithm to obtain this expression on quantum circuits with certainty. 
We devote Section \ref{Subsec:extract_certainty} for its discussion.

\subsection{Representation of time-evolution}\label{Subsec:time_evol_repr}

Here, we establish the way to express the time-evolution without Dyson-series expansion for the algorithm.
In an analogy to time-periodic cases in Ref. \cite{Mizuta2022optimal}, we aim at reproducing the time-evolution operator $U(t)$ by a unitary operation on the Floquet-Hilbert space.
Let $W_f^L$ denote a unitary gate which generates a uniform superposition as
\begin{equation}\label{Eq4:def_uniform_gate}
    W_f^L \ket{\vec{0}}_f = \frac{1}{\sqrt{(2L)^n}} \sum_{\vec{l} \in [L]^n} \ket{\vec{l}}_f,
\end{equation}
which yields at-most $\order{\log L}$ elementary gates.
For some natural numbers $p,q$ satisfying $p < q$, we define a unitary operation $\ms{U}_{p,q}^L(t)$ by
\begin{equation}\label{Eq4:def_extended_Ut}
    \ms{U}_{p,q}^L(t) = (W_f^{qL} \otimes I)^\dagger e^{-i \ms{H}_\mr{LP}^{qL} t} e^{-i \ms{H}^{qL} t} (W_f^{pL} \otimes I),
\end{equation}
which acts on the Floquet-Hilbert space $\mcl{H}_\mr{F}^{qL}$.
In this section, we prove that the time-evolution $U(t)$ is provided by this unitary gate as
\begin{equation}\label{Eq4:Approx_block_encode_Ut}
    \braket{\vec{0}|\ms{U}_{p,q}^L(t)|\vec{0}}_f \simeq \left( \frac{p}{q} \right)^{\frac{n}{2}} U(t),
\end{equation}
if the cutoff $L$ is sufficiently large.
In other words, $\ms{U}_{p,q}^L(t)$ is interpreted as an approximate block-encoding of $U(t)$.

\subsubsection{Preliminary results for proving Eq. (\ref{Eq4:Approx_block_encode_Ut})} From the definition of $\ms{U}_{p,q}^L(t)$ as Eq. (\ref{Eq4:def_extended_Ut}), the diagonal element of interest is computed as follows;
\begin{eqnarray}
    \braket{\vec{0}|\ms{U}_{p,q}^L(t)|\vec{0}}_f &=& \bra{\vec{0}}_f W_f^{qL \dagger} e^{-i \ms{H}_\mr{LP}^{qL} t} e^{-i \ms{H}^{qL} t} W_f^{pL} \ket{\vec{0}}_f \nonumber \\
    &=& \frac{1}{\sqrt{(2pL)^n (2qL)^n}} \sum_{\vec{l}^\prime \in [pL]^n} U_{\vec{l}^\prime}^{qL}(t), \label{Eq4:extended_extracted_Ut}\\
    U_{\vec{l}^\prime}^{qL}(t) &=& \sum_{\vec{l} \in [qL]^n} e^{-i \vec{l} \cdot \vec{\omega} t} \braket{\vec{l}|e^{-i \ms{H}^{qL} t}|\vec{l}^\prime}. \label{Eq4:def_extracted_Ut}
\end{eqnarray}
We first concentrate on evaluating $U_{\vec{l}^\prime}^{qL}(t)$ before proving Eq. (\ref{Eq4:Approx_block_encode_Ut}).
Remarkably, we can prove the following theorem on $U_{\vec{l}^\prime}^{qL}(t)$, which claims that it approximates the time-evolution operator $U(t)$ under sufficiently large $L$ regardless of $\vec{l}^\prime \in [pL]^n$.

\begin{theorem}\label{Thm:Floquet_Hilbert_repr}
\textbf{(Floquet-Hilbert space description)}

We impose all the assumptions in Section \ref{Subsec:Setup} on multi-periodic Hamiltonians $H(t)$, and assume $p<q$ for $p,q \in \bbN$ and $L \geq e^2 m_\imax \gamma t + m_\imax$ where $\gamma$ is defined by Eq. (\ref{Eq3:def_gamma}).
Then, for every point $\vec{l}^\prime \in [pL]^n$, the operator $U_{\vec{l}^\prime}^{qL}(t)$ [See Eq. (\ref{Eq4:def_extracted_Ut})] approximates the time-evolution operator $U(t)$ as
\begin{equation}\label{Eq4:Ut_error_bound_thm}
    \norm{U(t)-U_{\vec{l}^\prime}^{qL}(t)} \leq C_{n,m_\imax} \overline{\alpha}_0 t \left( \frac{e^2 m_\imax \gamma t}{L-m_\imax}\right)^{L/m_\imax-1},
\end{equation}
with $\overline{\alpha}_0 = \alpha - \alpha_{\vec{0}}$.
The $\order{1}$ constant $C_{n,m_\imax}$ is expressed by
\begin{equation}\label{Eq4:constant_nm}
    C_{n,m_\imax} = 4 (2 \sqrt{\pi} m_\imax)^n \frac{\Gamma(n)}{\Gamma(n/2)} e^{\sqrt{n}/m_\imax},
\end{equation}
where $\Gamma(x)$ denotes the gamma function.
\end{theorem}

\textit{Proof.---} Throughout this proof, we omit the subscript $f$ in $\ket{\vec{l}}_f$ and $\ket{\vec{l}^\prime}_f$.
We begin with differentiating $U^{qL}_{\vec{l}^\prime}(t)$ in time as follows,
\begin{eqnarray}
    && \dv{t} U_{\vec{l}^\prime}^{qL}(t) \nonumber \\
    && \quad = -i \sum_{\vec{l} \in [qL]^n}  e^{-i \vec{l} \cdot \vec{\omega} t} \braket{\vec{l}|(\vec{l}\cdot\vec{\omega} + \ms{H}^{qL}) e^{-i \ms{H}^{qL} t}|\vec{l}^\prime} \nonumber \\
    && \quad =  -i \sum_{\vec{l} \in [qL]^n} \sum_{\vec{m} \in M} e^{-i(\vec{l} \oplus \vec{m}) \cdot \vec{\omega} t} H_{\vec{m}} \bra{\vec{l}} e^{-i \ms{H}^{qL} t} \ket{\vec{l}^\prime}. \nonumber \\
    &&
\end{eqnarray}
In the second equality, $\vec{l} \oplus \vec{m}$ comes from the addition $\mr{Add}_{\vec{m}}^{qL}$ on the torus $[qL]^n$ as Eq. (\ref{Eq4:addition_torus}), and hence each $l_i \oplus m_i \in [qL]$ is defined modulo $2qL$ for each direction $i=1,2,\hdots,n$.
We substitute $-i H(t) U^{qL}_{\vec{l}^\prime}(t)$ from the above equality, which results in
\begin{eqnarray}
    && \dv{t} U^{qL}_{\vec{l}^\prime}(t) + i H(t) U^{qL}_{\vec{l}^\prime}(t) \nonumber \\
    && \quad = \sum_{\vec{l} \in [qL]^n} \sum_{\vec{m} \in M} f(\vec{l},\vec{m})  H_{\vec{m}} \bra{\vec{l}} e^{-i \ms{H}^{qL} t} \ket{\vec{l}^\prime} \equiv E(t), \nonumber \\
    && \label{Eq4:differential_eq_Ut}
\end{eqnarray}
where the factor $f(\vec{l},\vec{m})$ is defined by
\begin{equation}
    f(\vec{l},\vec{m}) = e^{-i(\vec{l} \oplus \vec{m}) \cdot \vec{\omega} t} - e^{-i(\vec{l} + \vec{m}) \cdot \vec{\omega} t}.
\end{equation}
We can solve the time-dependent differential equation Eq. (\ref{Eq4:differential_eq_Ut}) with using the initial condition $U^{qL}_{\vec{l}^\prime}(0)=I$ from Eq. (\ref{Eq4:def_extracted_Ut}).
This results in 
\begin{equation}
    U^{qL}_{\vec{l}^\prime}(t) = U(t) + \int_0^t \dd t^\prime U(t) U(t^\prime)^\dagger E(t^\prime), 
\end{equation}
and hence the error of interest is bounded by
\begin{equation}\label{Eq4:Ut_error_bound_Et}
    \norm{U(t)-U^{qL}_{\vec{l}^\prime}(t)} \leq t \sup_{t^\prime \in [0,t]} (\norm{E(t^\prime)}).
\end{equation}

We concentrate on evaluate the bound on $\norm{E(t)}$.
The factor $f(\vec{l},\vec{m})$ implies that each term in the summation of Eq. (\ref{Eq4:differential_eq_Ut}) can survive only when $\vec{l}$ is included in the boundary of $[qL]^n$, i.e., $\partial [qL]^n \equiv [qL]^n \backslash [qL-m_\imax]^n$.
We also note that the terms corresponding to $\vec{m}=\vec{0}$ always vanish by $\vec{l} \oplus \vec{m} = \vec{l} + \vec{m}$.
This results in the inequality,
\begin{equation}\label{Eq4:Et_bound_1}
    \norm{E(t)} \leq \sum_{\vec{m} \in M; \vec{m} \neq \vec{0}} 2 \norm{H_{\vec{m}}} \sum_{\vec{l} \in \partial [qL]^n} \norm{\bra{\vec{l}} e^{-i \ms{H}^{qL} t} \ket{\vec{l}^\prime}}.
\end{equation}
Then, we use the Lieb-Robinson bound in the Floquet-Hilbert space, which gives the upper bound of the transition amplitude $\braket{\vec{l}|e^{-i \ms{H}^{qL} t}|\vec{l}^\prime}$.
When we define the distance between $\vec{l}$ and $\vec{l}^\prime$ measured on the torus $[qL]^n$ by
\begin{equation}
    d^{qL}(\vec{l},\vec{l}^\prime) = \sqrt{\sum_{i=1}^n \left( \min \{ |l_i-l_i^\prime|, 2qL - |l_i-l_i^\prime| \} \right)^2},
\end{equation}
the transition amplitude is bounded from above by
 \begin{equation}\label{Eq4:Lieb_Robinson}
     \norm{\braket{\vec{l}|e^{-i \ms{H}^{qL} t}|\vec{l}^\prime}} \leq \left( \frac{e m_\imax \gamma t}{d^{qL}(\vec{l},\vec{l}^\prime)} \right)^{d^{qL}(\vec{l},\vec{l}^\prime)/m_\imax},
 \end{equation}
in case $d^{qL}(\vec{l},\vec{l}^\prime) \geq 2 m_\imax \gamma t$.
Since the derivation is similar to that for time-periodic Hamiltonians \cite{Mizuta2022optimal}, we provide it in Appendix \ref{Sec:Lieb_Robinson}.
For every point $\vec{l}^\prime \in [pL]^n$, the distance from $\vec{l} \in \partial [qL]^n$ satisfies
\begin{eqnarray}
    d^{qL}(\vec{l},\vec{l}^\prime) &\geq& (qL-m_\imax) - pL_\imax \nonumber \\
    &\geq& L-m_\imax \geq e^2 m_\imax \gamma t,
\end{eqnarray}
due to the assumptions, and hence we can safely apply the Lieb-Robinson bound, Eq. (\ref{Eq4:Lieb_Robinson}).
By combining Eq. (\ref{Eq4:Et_bound_1}) with the relation $\norm{H_{\vec{m}}} \leq \alpha_{\vec{m}}$, the error $\norm{E(t)}$ is further bounded by
\begin{eqnarray}
    \norm{E(t)} &\leq& 2 (\alpha-\alpha_{\vec{0}}) \sum_{\vec{l} \in \partial [qL]^n} \left( \frac{e m_\imax \gamma t}{d^{qL}(\vec{l},\vec{l}^\prime)} \right)^{d^{qL}(\vec{l},\vec{l}^\prime)/m_\imax} \nonumber \\
    &\leq& 2\overline{\alpha}_0 \left( \frac{e m_\imax \gamma t}{L-m_\imax} \right)^{(L-m_\imax)/m_\imax} \nonumber \\
    && \times \sum_{\vec{l} \in \partial [qL]^n} \left( \frac{e m_\imax \gamma t}{e^2 m_\imax \gamma t} \right)^{\frac{d^{qL}(\vec{l},\vec{l}^\prime)-(L-m_\imax)}{m_\imax}} \nonumber \\
    &\leq& 2\overline{\alpha}_0 \left( \frac{e^2 m_\imax \gamma t}{L-m_\imax} \right)^{\frac{L-m_\imax}{m_\imax}} \sum_{\vec{l} \in \partial [qL]^n} e^{-\frac{d^{qL}(\vec{l},\vec{l}^\prime)}{m_\imax}}. \nonumber \\
    \label{Eq4:Et_bound_2} &&
\end{eqnarray}

The remaining task for evaluating $\norm{E(t)}$ is to compute the summation over $\vec{l} \in \partial [qL]^n$ in Eq. (\ref{Eq4:Et_bound_2}).
Let $D_{\vec{l}^\prime}$ be a closed orthant to which $\vec{l}^\prime$ belongs, represented by $D_{\vec{l}^\prime} = \{ \vec{x} \in \bbR^n \, ; \, x_i \mr{sgn}(l_i^\prime) \geq 0 \, (i=1,2,\hdots,n) \}$ with $\mr{sgn}(0)\equiv 0$.
The summation over $\vec{l}^\prime$ in Eq. (\ref{Eq4:Et_bound_2}) is bounded by
\begin{eqnarray}
    \sum_{\vec{l} \in \partial [qL]^n} e^{-\frac{d^{qL}(\vec{l},\vec{l}^\prime)}{m_\imax}} &\leq& 2^n \sum_{\vec{l} \in \partial [qL]^n \cap D_{\vec{l}^\prime}} e^{-\frac{|\vec{l}-\vec{l}^\prime|}{m_\imax}} \nonumber \\
    &\leq& 2^n \int_{L-m_\imax}^\infty \dd r S_n r^{n-1} e^{-(r-\sqrt{n})/m_\imax} \nonumber \\
    &\leq& (2 m_\imax)^n e^{\sqrt{n}/m_\imax} S_n \Gamma (n), \label{Eq4:Et_summation_bound}
\end{eqnarray}
where $S_n \equiv 2 \pi^{n/2} / \Gamma (n/2)$ gives the surface area of an $n$-dimensional unit ball.
Finally, with using the relations Eqs. (\ref{Eq4:Ut_error_bound_Et}), (\ref{Eq4:Et_bound_2}), and (\ref{Eq4:Et_summation_bound}), we arrive at the bound on $\norm{U(t)-U_{\vec{l}^\prime}^{qL}(t)}$ in the form of Eq. (\ref{Eq4:Ut_error_bound_thm}), which completes the proof. $\quad \square$

Since the right hand side of Eq. (\ref{Eq4:Ut_error_bound_thm}) vanishes under $L \to \infty$, Theorem \ref{Thm:Floquet_Hilbert_repr} says that we can reproduce the exact time-evolution operator $U(t)$ by $U_{\vec{l}^\prime}^{qL}(t)$ with sufficiently large $L$.

\subsubsection{Proof of Eq. (\ref{Eq4:Approx_block_encode_Ut})}

We are ready to prove Eq. (\ref{Eq4:Approx_block_encode_Ut}), which indicates that the unitary operator $\ms{U}_{p,q}^L(t)$ can provide block-encoding of the time-evolution $U(t)$.
This relation is precisely stated by the following theorem.

\begin{theorem}\label{Thm:Block_encode_Ut}
\textbf{(Block-encoding of time-evolution)}

We impose all the assumptions in Section \ref{Subsec:Setup} on multi-periodic Hamiltonians $H(t)$, and assume $p < q$ for $p,q \in \bbN$.
When we choose the cutoff $L$ for the Fourier indices by
\begin{equation}\label{Eq4:Scaling_cutoff}
    L \in \order{ \alpha t + \frac{\log (1/\varepsilon)}{\log (e+(\alpha t)^{-1} \log (1/\varepsilon))}},
\end{equation}
the unitary gate $\ms{U}_{p,q}^L(t)$ on the Floquet-Hilbert space $\mcl{H}_\mr{F}^{qL}$ can generate the block-encoding of the time-evolution $U(t)$ as
\begin{equation}\label{Eq4:Thm_block_encode_Ut}
    \norm{\braket{\vec{0}|\ms{U}_{p,q}^L(t)|\vec{0}}_f - \left(\frac{p}{q}\right)^{\frac{n}{2}} U(t)} \leq \left( \frac{p}{q} \right)^{\frac{n}{2}}\varepsilon.
\end{equation}
\end{theorem}

\textit{Proof of Theorem \ref{Thm:Block_encode_Ut}.---} We choose the cutoff $L$ so that the difference $\norm{U(t)-U_{\vec{l}^\prime}^{qL}(t)}$ becomes smaller than the desirable error $\varepsilon$.
We apply the inequality Eq. (\ref{Eq2:ineq_Lambert_W}) with $\kappa = e^2 \gamma t$, $\eta = \varepsilon (C_{n,m_\imax} \overline{\alpha}_0 t)^{-1}$, and $x=L/m_\imax-1$, and set $L$ by
\begin{eqnarray}
    L &=& m_\imax \left\lceil e^3 \gamma t + \frac{4 \log (\frac{C_{n,m_\imax} \overline{\alpha}_0 t}{\varepsilon})}{\log (e+\frac{1}{e^2 \gamma t}\log (\frac{C_{n,m_\imax} \overline{\alpha}_0 t}{\varepsilon})) } + 1 \right\rceil \nonumber \\
    && \label{Eq4:Explicit_cutoff} \\
    &\leq& e^3 m_\imax \gamma t  + \frac{4 m_\imax \log (1/\varepsilon)}{\log (e+(e^2 \gamma t)^{-1} \log (1/\varepsilon))} \nonumber \\
    && \qquad + 4 m_\imax \log ( C_{n,m_\imax} \overline{\alpha}_0 t) + m_\imax + 1.
\end{eqnarray}
Then, Theorem \ref{Thm:Floquet_Hilbert_repr} ensures $\norm{U(t)-U_{\vec{l}^\prime}^{qL}(t)} \leq \varepsilon$.
We also note that the above choice of $L$ satisfies the scaling of Eq. (\ref{Eq4:Scaling_cutoff}) since we have the assumption $n,m_\imax \in \order{1}$ and the relation $\gamma, \overline{\alpha}_0 \leq \alpha$ from Eqs. (\ref{Eq3:def_gamma}) and (\ref{Eq3:ineq_alpha}).
The expression of $\ms{U}_{p,q}^L(t)$ given by Eq. (\ref{Eq4:extended_extracted_Ut}) leads to
\begin{eqnarray}
    && \norm{\braket{\vec{0}|\ms{U}_{p,q}^L(t)|\vec{0}}_f - \left(\frac{p}{q}\right)^{\frac{n}{2}} U(t)} \nonumber \\  
    && \quad\leq \frac{1}{\sqrt{(2pL)^n (2qL)^n}} \sum_{\vec{l}^\prime \in [pL]^n} \norm{U(t)-U_{\vec{l}^\prime}^{qL}(t)} \nonumber \\ 
    && \quad \leq \left(\frac{p}{q}\right)^{\frac{n}{2}} \varepsilon,
\end{eqnarray}
which completes the proof of Eq. (\ref{Eq4:Thm_block_encode_Ut}). $\quad \square$

\subsection{Extracting time-evolution with certainty}\label{Subsec:extract_certainty}
 
Theorem \ref{Thm:Block_encode_Ut} claims that the unitary operator $\ms{U}_{p,q}^L(t)$ gives the block-encoding of the time-evolution $U(t)$.
We next provide the way to exploit it for accurately obtaining the time-evolved state $U(t) \ket{\psi}$ with certainty, using the oblivious amplitude amplification \cite{Berry2017-oblivious}.
With defining the projections $\ms{P}_0 \equiv \ket{\vec{0}}\bra{\vec{0}}_f \otimes I$ and $\ms{P}_0^\perp = I - \ms{P}_0$, the block-encoding provides the time-evolution probabilistically by
\begin{eqnarray}
    \ms{U}_{p,q}^{L}(t) \ket{\vec{0}}_f \ket{\psi} &=& (\ms{P}_0 + \ms{P}_0^\perp) \ms{U}_{p,q}^{L}(t) \ket{\vec{0}}_f \ket{\psi} \nonumber \\
    &=& \ket{\vec{0}}_f \braket{\vec{0}|\ms{U}_{p,q}^L(t)|\vec{0}}_f \ket{\psi} + \ket{\Psi^\perp} \nonumber \\
    &=& \left(\frac{p}{q}\right)^{\frac{n}{2}} \ket{\vec{0}}_f \left\{ U(t) \ket{\psi} + \ket{\psi_\varepsilon} \right\} + \ket{\Psi^\perp}, \nonumber \\
    &&
\end{eqnarray}
where $\ket{\Psi^\perp} \equiv \ms{P}_0^\perp \ms{U}_{p,q}^{L}(t) \ket{\vec{0}}_f$ is always orthogonal to $\ket{\vec{0}}_f$.
The deviation $\ket{\psi_\varepsilon}$, which is given by
\begin{equation}
    \ket{\psi_\varepsilon} \equiv \left( \frac{q}{p} \right)^{\frac{n}{2}} \braket{\vec{0}|\ms{U}_{p,q}^L(t)|\vec{0}}_f \ket{\psi} - U(t) \ket{\psi},
\end{equation}
has the norm bounded by $\norm{\ket{\psi_\varepsilon}} \leq \varepsilon$ by Theorem \ref{Thm:Block_encode_Ut}.
When we measure the ancilla system $f$ and post-select the result $\ket{\vec{0}}_f$, we can obtain the target state $U(t) \ket{\psi}$ with the allowable deviation $\ket{\psi_\varepsilon}$.
The success probability of post-selection $P_\mr{success}$ amounts to
\begin{eqnarray}
    P_\mr{success} &=& \norm{\braket{\vec{0}|\ms{U}_{p,q}^L(t)|\vec{0}}_f \ket{\psi}}^2 \nonumber \\
    &=& \left(\frac{p}{q}\right)^n (1+ 2 \Re{\braket{\psi_\varepsilon|U(t)|\psi}} + \braket{\psi_\varepsilon|\psi_\varepsilon}). \nonumber \\
    &&
\end{eqnarray}
Considering the relation $|\braket{\psi_\varepsilon|U(t)|\psi}| \leq \varepsilon$ and $\braket{\psi_\varepsilon|\psi_\varepsilon} \leq \varepsilon^2 < \varepsilon$ for $\varepsilon \in [0,1]$, the success probability lies in
\begin{equation}
    P_\mr{success} \in \left[ \left(\frac{p}{q}\right)^n (1-3\varepsilon), \left(\frac{p}{q}\right)^n (1+3\varepsilon) \right].
\end{equation}
The condition $p < q$ ($p,q \in \bbN$) suggests that the unitary gate $\ms{U}_{p,q}^L(t)$ fails to achieve success probability sufficiently close to $1$.

In order to deterministically get the time-evolved state $\ket{\psi}$, we employ the oblivious amplitude amplification which amplifies the success probability \cite{Berry2017-oblivious}.
With a phase rotation operator $\ms{R}(\varphi)=e^{i\varphi (2\ket{\vec{0}}\bra{\vec{0}}_f - I)} \otimes I$, it exploits a unitary operator defined by
\begin{equation}\label{Eq4:def_amplification}
    \ms{U}^L (t) = e^{i\varphi_0} \left\{ \ms{U}_{p,q}^L(t) \ms{R}(\varphi_1) [\ms{U}_{p,q}^L(t)]^\dagger \ms{R}(\varphi_2) \right\}^D \ms{U}_{p,q}^L(t),
\end{equation}
with tunable parameters $\varphi_0,\varphi_1,\varphi_2 \in \bbR$ and an iteration number $D \in \bbN$.
In a similar manner to the generalized Grover's search algorithm with flexible rotation angles $\varphi_1,\varphi_2$ \cite{Grover1997-prl,Hoyer2000-grover,Long2001-grover}, 
we define the iteration number by
\begin{equation}
    D \geq \frac{\pi}{4 \arcsin (\sqrt{P})} - \frac{1}{2} \in \order{\frac{1}{\sqrt{P}}}
\end{equation}
when the success probability satisfies $P_\mr{success} \geq P - \order{\varepsilon}$.
Then, there exists a parameter set $\varphi_0,\varphi_1,\varphi_2 \in \bbR$ realizing the success probability $1$ as
\begin{equation}
    \ms{U}^L(t) \ket{\vec{0}}_f \ket{\psi} = \ket{\vec{0}}_f U(t) \ket{\psi} + \order{D\varepsilon}, \quad ^\forall \ket{\psi} \in \mcl{H}.
\end{equation}
We set the natural numbers $p$ and $q$ by
\begin{equation}
    p = n, \quad q= n+1,
\end{equation}
with the number of the frequencies $n \in \order{1}$.
This choice ensures the lower bound of the success probability
\begin{eqnarray}
    P_\mr{success} \geq \left( 1 + \frac{1}{n} \right)^{-n} (1-3\varepsilon) \geq \frac{1}{e}(1 - 3 \varepsilon ).
\end{eqnarray}
The choice of the iteration number by
\begin{equation}
    D = \left\lceil \frac{\pi}{4 \arcsin (\sqrt{e^{-1}})} - \frac12 \right\rceil = 1
\end{equation}
is enough in this case.
Based on the generalized Grover's search algorithm by Ref. \cite{Long2001-grover}, the rotation phases $\varphi_1, \varphi_2$ are chosen by
\begin{eqnarray}
    \varphi_1 = \varphi_2 &=& \arcsin \left\{ \frac{1}{\sqrt{P}} \sin (\frac{\pi}{4D+2}) \right\} \nonumber \\
    &=& \arcsin \left\{ \frac12 \left( 1+n^{-1} \right)^{n/2} \right\},
\end{eqnarray}
while the global phase $\varphi_0$ is irrelevant in practice.

In consequence, the unitary operator $\ms{U}^L(t)$ on the Floquet-Hilbert space $\mcl{H}_\mr{F}^{(n+1)L}$ generates the target time-evolved state $U(t) \ket{\psi}$ with an error up to $\order{\varepsilon}$.
According to Eqs. (\ref{Eq4:def_extended_Ut}) and (\ref{Eq4:def_amplification}), the computational resources required for $\ms{U}^L(t)$ is summarized as follows;
\begin{enumerate}[(i)]
    \item Ancilla system $\{ \ket{\vec{l}} \}_{\vec{l} \in [(n+1)L]^n}$ for labeling Fourier indices.
    It requires $\lceil \log_2 \{(n+1)L\}^n \rceil \in \order{\log L}$ qubits.
   
   \item $2D+1 = 3$ times usage of the time-evolution operators $e^{-i \ms{H}_\mr{LP}^{(n+1)L} t}$ and $e^{-i \ms{H}^{(n+1)L} t}$, or their inverse operations.
   
   \item The other gates on the ancilla system $f$, i.e., the unitary gates $W_f^{nL}$, $W_f^{(n+1)L}$ [See Eq. (\ref{Eq4:def_uniform_gate})], and $\ms{R}(\varphi)=e^{i \varphi (2\ket{\vec{0}}\bra{\vec{0}}_f -I)} \otimes I$.
   Each of them is used $\order{1}$ times, requiring $\order{\log L}$ elementary gates.
   
\end{enumerate}
In terms of the number of quantum gates, the dominant task in the Hamiltonian simulation is to implement the time-evolution under the time-independent Hamiltonians $\ms{H}_\mr{LP}^{(n+1)L}$ and $\ms{H}^{(n+1)L}$.
We use the qubitization technique \cite{Low2019-qubitization}, introduced in Section \ref{Sec:Preliminaries}, for this implementation in the following section.

\section{Time-evolution operators under effective Hamiltonian}\label{Sec:Qubitization_eff}

The unitary gate $\ms{U}^L(t)$, given by Eq. (\ref{Eq4:def_amplification}), executes Hamiltonian simulation under multi-periodic Hamiltonians, with making use of $\order{1}$ times queries to the time-evolution operators under the Hamiltonians on the Floquet-Hilbert space, $\ms{H}_\mr{LP}^{(n+1)L}$ and $\ms{H}^{(n+1)L}$.
Since they are no longer time-dependent, we can employ the QET-based qubitization technique for implementing the time-evolution operators under them.
In this section, we identify how much resource is required for the qubitization in terms of the supposed oracles $\{ O_{\vec{m}} \}_{\vec{m} \in M}$ (block-encoding of each Fourier component $H_{\vec{m}}$) and some other elementary gates.

\subsection{Linear potential Hamiltonian}\label{Subsec:block_encode_linear_potential}

We first evaluate the cost for implementing the time-evolution operator under the linear potential Hamiltonian $\ms{H}_\mr{LP}^{(n+1)L}$, given by Eq. (\ref{Eq4:def_H_linear_potential}).
For brevity, we organize the block-encoding of $\ms{H}_\mr{LP}^L$ with omitting the $\order{1}$ coefficient in the cutoff, $n+1$.
Each ancilla state $\ket{\vec{l}}_f$ for $\vec{l} \in [L]^n$ can be decomposed into $\ket{\vec{l}}_f = \bigotimes_{i=1}^n \ket{l_i}_{f_i}$ with $l_i \in [L]$, and then the linear potential Hamiltonian is rewritten by
\begin{equation}
    \ms{H}_\mr{LP}^L = \sum_{i=1}^n \left( \sum_{l_i \in [L]}l_i \omega_i \ket{l_i}\bra{l_i}_{f_i} \right) \otimes I_{f \backslash f_i} \otimes I.
\end{equation}
Thus, it is a linear combination of one-dimensional linear potential Hamiltonians
Each of them has an efficiently-implementable block-encoding $O_\mr{LP}^i$ \cite{Mizuta2022optimal}, such that
\begin{equation}
    \braket{0|O_\mr{LP}^i|0}_{d} = \frac{1}{2L \omega_i} \sum_{l_i \in [L]}l_i \omega_i \ket{l_i}\bra{l_i}_{f_i},
\end{equation}
which can be equipped with $\order{\log L}$ elementary gates and an $\order{\log L}$-qubit ancilla trivial state $\ket{0}_{d}$.
Once block-encoding of each term in a linear combination is obtained, its block-encoding is also immediately organized \cite{Gilyen2019-qsvt}.
We use the ancilla system $c$ having $n$ levels and the oracle $G_\mr{freq}$, defined by Eq. (\ref{Eq3:def_G_freq}).
The block-encoding for $\ms{H}_\mr{LP}^L$ is organized by
\begin{equation}\label{Eq5:oracle_linear_potential}
    \ms{O}_\mr{LP}^L = G_\mr{freq}^\dagger \left( \sum_{i=0}^{n-1} \ket{0}\bra{0}_c \otimes O_\mr{LP}^i \right) G_\mr{freq},
\end{equation}
where we omit identity operators. 
We can confirm the relation,
\begin{equation}
    \braket{0|\ms{O}_\mr{LP}|0}_{c d} = \sum_{i=1}^n \frac{\omega_i}{\omega} \braket{0|O_\mr{LP}^i|0}_{d} = \frac{\ms{H}_\mr{LP}^L}{2L\omega}.
\end{equation}

We employ the qubitization technique to implement $\exp (-i \ms{H}_\mr{LP}^L t)$ with using the block-encoding $\ms{O}_\mr{LP}^L$.
The unitary gate $\ms{O}_\mr{LP}^L$ requires two queries to the supposed oracle $G_\mr{freq}$.
As a results, the qubitization technique dictates the existence of a unitary gate $\ms{U}_\mr{LP}^L(t)$ such that
\begin{equation}
    \ms{U}_\mr{LP}^L(t) \ket{0}_{cd} \ket{\Psi} = \ket{0}_{cd} e^{-i \ms{H}_\mr{LP}^L t} \ket{\Psi} + \order{\varepsilon},
\end{equation}
($^\forall \ket{\Psi} \in \mcl{H}_\mr{F}^{L}$), which can be implemented with following resources;
\begin{itemize}
    \item Ancilla qubits for $c$ and $d$;
    \begin{equation}
        \lceil \log_2 n \rceil + \order{\log L} \in \order{\log L}.
    \end{equation}
    \item Query complexity of the oracle $G_\mr{freq}$;
    \begin{equation}\label{Eq5:Query_complexity_linear_potential}
        \order{L\omega t + \frac{\log (1/\varepsilon)}{\log (e+(L\omega t)^{-1} \log (1/\varepsilon))}}
    \end{equation}
    \item Additional elementary gates per query; $\order{\log L}$
\end{itemize}
The actual algorithm works with the time-evolution $\exp (-i \ms{H}_\mr{LP}^{(n+1)L} t)$, and the accurate cost is obtained by substituting $(n+1)L$ into $L$ in the above results.
Since $n$ is supposed to be an $\order{1}$ constant, the resulting cost is essentially the same as the above one.

\subsection{Effective Hamiltonian}

We discuss the cost for implementing the time-evolution $\exp (-i \ms{H}^{(n+1)L} t)$.
Here, we again omit the coefficient $n+1$ and consider the block-encoding of the effective Hamiltonian $\ms{H}^L$.
We concentrate on the first term in Eq. (\ref{Eq4:def_H_eff}), 
\begin{equation}
    \ms{H}_\mr{Add}^L \equiv \sum_{\vec{m} \in M} \mr{Add}_{\vec{m}}^L \otimes H_{\vec{m}}. 
\end{equation}
The block-encoding of $\ms{H}_\mr{Add}^L$ can be composed of the oracles $O_{\vec{m}}$ and $G_\mr{coef}$ [See Section \ref{Subsec:Main_results}] by
\begin{equation}\label{Eq5:oracle_add_term}
    \ms{O}_\mr{Add}^L = G_\mr{coef}^\dagger \left( \sum_{\vec{m} \in M} \ket{\vec{m}}\bra{\vec{m}}_b \otimes O_{\vec{m}} \right) G_\mr{coef},
\end{equation}
where we need the $\order{\log (|M|)}$-qubit ancilla system $b$.
It satisfies
\begin{equation}
    \braket{0|\ms{O}_\mr{Add}^L|0}_{ab} = \sum_{\vec{m} \in M} \frac{\alpha_{\vec{m}}}{\alpha} \braket{0|O_{\vec{m}}|0}_a = \frac{\ms{H}_\mr{Add}^L}{\alpha},
\end{equation}
where we define the reference state by $\ket{0}_{ab} = \ket{0}_{a} \otimes \ket{\vec{0}}_b$.

We introduce another ancilla qubit $e$ to organize the block-encoding of $\ms{H}^L$.
Under the preparation of the ancilla system $a^\prime$ composed of the systems $a,b,c,d,e$, and $f$, we define a unitary gate $\ms{O}^L$ on it by
\begin{equation}
    \ms{O}^L =  \ms{R}^L \left( \ket{0}\bra{0}_e \otimes \ms{O}_\mr{Add}^L - \ket{1}\bra{1}_e \otimes \ms{O}_\mr{LP}^L \right) \ms{R}^L,
\end{equation}
where the unitary gate $\ms{R}^L$ denotes a single-qubit rotation around $Y$-axis on the system $e$ as
\begin{equation}
    \ms{R}^L = \left( e^{ - i \theta_L Y} \right)_e, \quad \theta_L = \arccos \left( \sqrt{\frac{\alpha}{\alpha + 2L\omega}}\right).
\end{equation}
The unitary operator $\ms{O}_L$ provides the block-encoding of $\ms{H}^L$ with the reference state $\ket{0}_{a^\prime} = \ket{0}_a \ket{\vec{0}}_b \ket{0}_c \ket{0}_d \ket{0}_e$ as 
\begin{eqnarray}
    \braket{0|\ms{O}^L|0}_{a^\prime} &=& \frac{\alpha \braket{0|\ms{O}_\mr{Add}^L|0}_{ab} - 2L\omega \braket{0|O_\mr{LP}^i|0}_{d} }{\alpha+2L\omega} \nonumber \\
    &=& \frac{\ms{H}^L}{\alpha + 2L\omega}.
\end{eqnarray}
We can obtain the cost for implementing the time-evolution $\exp (-i\ms{H}^L t)$ with an allowable error $\varepsilon$ based on Section \ref{Sec:Preliminaries}.
The numbers of calls for the oracles $O_{\vec{m}}$, $G_\mr{coef}$, and $G_\mr{freq}$ are respectively given by $|M|$, $2$, and $2$ via Eqs. (\ref{Eq5:oracle_linear_potential}) and (\ref{Eq5:oracle_add_term}), and all of them can be attributed to $\order{1}$ numbers.
We can organize a unitary gate $\ms{U}_\eff^L(t)$ such that
\begin{equation}
    \ms{U}_\eff^L(t) \ket{0}_{a^\prime} \ket{\Psi} = \ket{0}_{a^\prime} e^{-i \ms{H}^L t} \ket{\Psi} + \order{\varepsilon},
\end{equation}
for arbitrary quantum states $\ket{\Psi} \in \mcl{H}_\mr{F}^L$ by the following resources.
\begin{itemize}
    \item Ancilla qubits for $a^\prime$; $n_a + \order{\log L}$.
    \item Query complexity counted by the oracles;
    \begin{equation}\label{Eq5:Query_complexity_H_eff}
        \order{(\alpha+L\omega)t + \frac{\log (1/\omega t)}{\log (e+(\alpha+L\omega)^{-1} \log (1/\varepsilon))}}.
    \end{equation}
    \item Additional gates per query; $\order{\log L}$.
\end{itemize}

\section{Algorithm of Hamiltonian simulation}\label{Sec:Algorithm}

In this section, we complete the algorithm for Hamiltonian simulation of multi-periodic time-dependent systems, and derive its computational cost.
Depending on the time scale of interest, we establish two different approaches.
The first case is $\order{1}$-period dynamics in which $\omega t \in \order{1}$ is satisfied, while the other is $\Omega(1)$-period dynamics with $\omega t \in \Omega (1)$.
Combining the explicit formula for the cutoff $L$, Eq. (\ref{Eq4:Explicit_cutoff}), and the query complexity, Eqs. (\ref{Eq5:Query_complexity_linear_potential}) and (\ref{Eq5:Query_complexity_H_eff}), a naive execution of the algorithm experiences quadratic increase of the cost in time $t$.
We need proof by cases based on the time scale to avoid such a problem.
We also note that the results provided below completely includes those for Hamiltonian simulation of time-periodic systems \cite{Mizuta2022optimal}, and share the same scaling of the cost.
In other words, we succeed in clarifying broader classes of time-dependent Hamiltonians that can be simulated as efficiently as time-independent Hamiltonians.

\subsection{$\order{1}$-period dynamics}\label{Subsec:algorithm_Order_1}

We here provide the algorithm for $\order{1}$-period dynamics.
This case is useful when we are interested in slow modulation in time with small $\omega$, exemplified by (quasi-)adiabatic quantum dynamics \cite{Albash2018RevModPhys,Nathan2021-multi}.

We first prepare the target system initialized to $\ket{\psi} \in \mcl{H}$ and the ancilla quantum system $f$ for Fourier indices, so that they can embody the Floquet-Hilbert space $\mcl{H}_\mr{F}^{(n+1)L}$.
Here, the cutoff $L$ is chosen by Eq. (\ref{Eq4:Explicit_cutoff}), whose scaling is given by Eq. (\ref{Eq4:Scaling_cutoff}).
With attaching additional ancilla qubits for $a$, $b$, $c$, $d$, and $e$ required for the qubitization technique, we initialize the combined ancilla system $a^\prime$ to the reference state $\ket{0}_{a^\prime}$.
We organize the unitary operator $\ms{U}^L(t)$ designated by Eq. (\ref{Eq4:def_amplification}) in which we substitute the results of qubitization, $\ms{U}_\mr{LP}^{(n+1)L}(t)$ and $\ms{U}_\eff^{(n+1)L}(t)$, instead of the time-evolution operators under $\ms{H}_\mr{LP}^{(n+1)L}$ and $\ms{H}^{(n+1)L}$.
The resulting unitary gate $\ms{U}^L(t)$ reproduces the time-evolution operator $U(t)$ as
\begin{equation}
    \ms{U}^L(t) \ket{0}_{a^\prime} \ket{\vec{0}}_f \ket{\psi} = \ket{0}_{a^\prime} \ket{\vec{0}}_f U(t) \ket{\psi} + \order{\varepsilon},
\end{equation}
which completes the algorithm.

Let us evaluate the computational cost.
The ancilla systems $a^\prime$ and $f$ respectively yield $n_a+\order{\log L}$ and $\order{\log L}$ qubits, whose scaling is designated by Eq. (\ref{Eq4:Scaling_cutoff}) as 
\begin{eqnarray}
    &&\order{\log L} \subset \nonumber \\
    && \quad \order{\log (\alpha t) + \log \log (1/\varepsilon) + \log  (m_\imax \log C_{n,m_\imax} )}. \nonumber \\
    &&
\end{eqnarray}
Although the constant $C_{n,m_\imax}$ super-exponentially increases in $n$ by Eq. (\ref{Eq4:constant_nm}), it hardly affects the number of ancilla qubits as
\begin{equation}
    \log (m_\imax \log C_{n,m_\imax}) \in \order{\log n + \log m_\imax},
\end{equation}
under the assumption $n, m_\imax \in \order{1}$.
Next, the query complexity $Q$, measured by the oracles $O_{\vec{m}}$, $G_\mr{coef}$, and $G_\mr{freq}$ is determined by the qubitization technique in Section \ref{Sec:Preliminaries}.
It is obtained by substituting the form of $L$, Eq. (\ref{Eq4:Scaling_cutoff}), into Eqs. (\ref{Eq5:Query_complexity_linear_potential}) and (\ref{Eq5:Query_complexity_H_eff}).
The oblivious amplitude amplification employed in Section \ref{Subsec:extract_certainty} affects the query complexity only by the multiplicative factor $\order{1}$.
Finally, we need additional gates other than the oracles, only acting on the ancilla systems.
They are composed of $\order{1}$-times use of $W_f^L$ [See Eq. (\ref{Eq4:def_uniform_gate})], $\order{1}$-times use of the phase rotation $\ms{R}(\varphi)$ for the oblivious amplitude amplification, and some other gates for the qubitization.
The last one has a dominant scaling in the total number of elementary gates, given by $\order{Q(n_a+\log L)}$, while those for the former two requires at-most $\order{\log L}$.
As a result, we arrive at the following theorem on Hamiltonian simulation of multi-periodic time-dependent systems.

\begin{theorem}\label{Thm:algorithm_Order1}
\textbf{(Cost for $\order{1}$-period dynamics)}

We impose all the assumptions in Section \ref{Subsec:Setup} on multi-periodic time-dependent Hamiltonians $H(t)$, and are allowed to use the oracles $O_{\vec{m}}$, $G_\mr{coef}$, and $G_\mr{freq}$.
The time of interest is supposed to be $t \in \order{1} \times T$.
Then, we can simulate the time-evolved state $U(t)\ket{\psi}$ from arbitrary initial states $\ket{\psi} \in \mcl{H}$ with a desirable error up to $\order{\varepsilon}$, by the following resources;
\begin{itemize}
    \item Number of ancilla qubits;
    \begin{equation}
        n_a + \order{\log (\alpha t) + \log \log (1/\varepsilon)}.
    \end{equation}
    \item Scaling of query complexity $Q$;
    \begin{equation}\label{Eq6:Query_Order_1}
        \alpha t + \frac{\log (1/\varepsilon)}{\log (e+\{ \alpha t + o(\log (1/\varepsilon))\}^{-1} \log (1/\varepsilon))}. 
    \end{equation}
    \item Additional gates per query;
    \begin{equation}
        \order{n_a + \log (\alpha t) + \log \log (1/\varepsilon)}.
    \end{equation}
\end{itemize}
In the query complexity, the term $o (\log (1/\varepsilon))$ exactly scales as
\begin{equation}
    \frac{\log (1/\varepsilon)}{\log (e+ (\alpha t)^{-1} \log (1/\varepsilon))},
\end{equation}
which is smaller than $\log(1/\varepsilon)$.
\end{theorem}

Let us discuss the efficiency of the algorithm and compare it with those of conventional algorithms.
When independently increasing the time $t$ or the inverse error $1/\varepsilon$ with fixing the other, we observe that the query complexity $Q$ has optimal linear scaling in $t$, and nearly-optimal scaling in $1/\varepsilon$ given by
\begin{equation}
    \order{\frac{\log (1/\varepsilon)}{\log \log \log (1/\varepsilon)}}.
\end{equation}
When we consider scaling both in $t$ and $1/\varepsilon$, we emphasize that our algorithm achieves additive scaling given by 
\begin{equation}
    \poly{N} t + o(\log(1/\varepsilon)).
\end{equation}
This is sufficiently close to the one for qubitization [See Eq. (\ref{Eq2:query_qubitization})], and saves much cost compared to the truncated Dyson series algorithm \cite{Low2018-dyson,Kieferova2019-dyson}, which gives productive scaling as
\begin{equation}\label{Eq6:Query_Dyson}
    \alpha t \frac{\log (\alpha t)}{\log \log (\alpha t / \varepsilon)}.
\end{equation}
In particular, when we impose $\alpha t \gtrsim \log (1/\varepsilon)$, which naturally arises from $\alpha \in \mr{poly}(N)$, Eq. (\ref{Eq6:Query_Order_1}) says that the query complexity of our algorithm corresponds to that of qubitization, which is the best one for time-independent cases.

\subsection{$\Omega (1)$-period dynamics}\label{Subsec:algorithm_Omega_1}

We establish the algorithm when the time $t$ is given by $\omega t \in \Omega (1)$.
It is useful in case we are interested in long-time dynamics when the frequency $\omega$ is not so small.
For instance, when we simulate quantum materials under multiple lights \cite{Chu2004-we-multi,Martin2017-multi} or nonequilibrium phases of matters under time-quasiperiodic drive \cite{Crowley2019-multi,Else2020-multi}, we need their dynamics over multiple periods $t \in \Omega (1) \times T$.

Our strategy of the algorithm is dividing the time $t$ into $r \equiv \lceil \omega t \rceil$ parts and repeating time-evolution for the separated time $t/r \in \order{1} \times T$.
The time-evolution operator can be split into
\begin{eqnarray}
    U(t) &=& \prod_{s=0}^{r-1} U((s+1)t/r; st/r), \\
    U(t_1;t_2) &=& \mcl{T} \exp \left( -i \int_{t_1}^{t_2} \dd t H(t)\right).
\end{eqnarray}
First, due to the relation $t/(rT) \in \order{1}$, we can execute the time-evolution $U(t/r,0)=U(t/r)$ by the algorithm provided in Section \ref{Subsec:algorithm_Order_1}, with setting the desirable error $\order{\varepsilon/r}$.
To be precise, when we introduce the cutoff $L_T$ for Fourier indices by substituting $t/r$ and $\varepsilon/r$ into $t$ and $\varepsilon$ of Eq. (\ref{Eq4:Explicit_cutoff}), which results in
\begin{eqnarray}
    L_T \in \order{\frac{\alpha t}{r} + \frac{\log (r/\varepsilon)}{\log (e+(\alpha t/r)^{-1} \log (r/\varepsilon))}},
\end{eqnarray}
we can organize a unitary gate $\ms{U}^{L_T}_{s=0}$ such that
\begin{equation}
    \ms{U}^{L_T}_{s=0} \ket{0}_{a^\prime} \ket{\vec{0}}_f \ket{\psi} = \ket{0}_{a^\prime} \ket{\vec{0}}_f U(t/r;0) \ket{\psi} + \order{\varepsilon/r}.
\end{equation}
The cost for $\ms{U}^{L_T}_{s=0}$ is obtained by $t/r$ and $\varepsilon/r$ into $t$ and $\varepsilon$ in Theorem \ref{Thm:algorithm_Order1}.

We repeat this procedure for the time-evolution $U((s+1)t/r;st/r)$ from $s=0$ to $s=r-1$.
We should be careful of the change in the time origin 
When we implement the time-evolution for each time step $[st/r,(s+1)t/r]$, we execute Hamiltonian simulation from $t^\prime=0$ to $t^\prime=t/r$ with the following Hamiltonian;
\begin{equation}
    H_s(t^\prime) = \sum_{\vec{m} \in M} \left( H_{\vec{m}} e^{-i \vec{m} \cdot \vec{\omega} st/r}\right) e^{-i \vec{m} \cdot \vec{\omega} t^\prime}.
\end{equation}
Each Fourier component $H_{\vec{m}}$ is replaced by $H_{\vec{m}} e^{-i \vec{m} \cdot \vec{\omega} st/r}$.
As a result, it is required to substitute $O_{\vec{m}} e^{-i \vec{m} \cdot \vec{\omega} st/r}$ into the oracle $O_{\vec{m}}$ in the block-encoding $\ms{O}_\mr{Add}^L$ [See Eq. (\ref{Eq5:oracle_add_term})].
This can be executed by inserting the unitary operation $\exp (- i \sum_{\vec{m} \in M} (st/r) \ket{\vec{m}}\bra{\vec{m}})$ into $\ms{O}_\mr{Add}^L$, and its cost is obtained in a similar way to Section \ref{Subsec:block_encode_linear_potential}.
Since the resulting cost does not affect the scaling, implementing each time-evolution $U((s+1)t/r;st/r)$ can be executed with essentially the same cost as $U(t/r)$.
After repetition until $s=r-1$, we obtain the target time-evolved state as
\begin{equation}\label{Eq6:Ut_time_split}
    \prod_{s=0}^{r-1} \ms{U}_s^{L_T} \ket{0}_{a^\prime} \ket{\vec{0}}_f \ket{\psi} = \ket{0}_{a^\prime} \ket{\vec{0}}_f U(t) \ket{\psi} + \order{\varepsilon}.
\end{equation}

Let us evaluate the computational cost.
First, the number of ancilla qubits is determined by the ancilla systems $a^\prime$ and $f$ and the choice of the cutoff $L_T$.
We note that, while each time time-evolution $U((s+1)t/r;st/r)$ needs the ancilla systems, we need a single pair of $a^\prime$ and $f$.
This is because the ancilla system is initialized to $\ket{0}_{a^\prime}\ket{\vec{0}}_f$ at the end of each step by Eq. (\ref{Eq6:Ut_time_split}), and can be reused.
The query complexity counted by the oracles $O_{\vec{m}}$, $G_\mr{coef}$, and $G_\mr{freq}$ is dominated by $r$ times implementation of the time-evolution $U((s+1)t/r;st/r)$.
We can specify its scaling by multiplying the query complexity Eq. (\ref{Eq6:Query_Order_1}) by $r = \lceil \omega t \rceil \in \order{\omega t}$ under the substitution $t \to t/r$ and $\varepsilon \to \varepsilon/r$.
The number of additional elementary gates is given in a similar way.
Finally, we arrive at the following theorem on the cost of Hamiltonian simulation for $\Omega (1)$-period dynamics.

\begin{theorem}\label{Thm:algorithm_Omega1}

\textbf{(Cost for $\Omega (1)$-period dynamics)}

We impose all the assumptions in Section \ref{Subsec:Setup} on multi-periodic time-dependent Hamiltonians $H(t)$, and are allowed to use the oracles $O_{\vec{m}}$, $G_\mr{coef}$, and $G_\mr{freq}$.
The time of interest is supposed to be $t \in \Omega (1) \times T$.
Then, we can simulate the time-evolved state $U(t)\ket{\psi}$ from arbitrary initial states $\ket{\psi} \in \mcl{H}$ with a desirable error up to $\order{\varepsilon}$, by the following resources;

\begin{itemize}
    \item Number of ancilla qubits;
    \begin{equation}
        n_a + \order{\log (\alpha / \omega) + \log \log (\omega t/\varepsilon)}.
    \end{equation}
    \item Scaling of query complexity $Q$;
    \begin{equation}\label{Eq6:Query_Omega_1}
        \alpha t +  \frac{\omega t \log (\omega t/\varepsilon)}{\log (e+(\alpha / \omega + o(\log (\omega t / \varepsilon)))^{-1} \log (\omega t/\varepsilon))}.
    \end{equation}
    
    \item Additional gates per query;
    \begin{equation}
        \order{n_a + \log (\alpha / \omega) + \log \log (\omega t/\varepsilon)}.
    \end{equation}
\end{itemize}
In the query complexity, the $o(\log (\omega t/\varepsilon))$ term scales as
\begin{equation}
    \frac{\log (\omega t / \varepsilon)}{\log (e+ (\alpha/\omega)^{-1} \log (\omega t/\varepsilon))}.
\end{equation}
\end{theorem}

Let us investigate the optimality of the query complexity and compare it with other algorithms presented in Table \ref{Table:comparison_algorithms}.
The scaling of Eq. (\ref{Eq6:Query_Omega_1}) in time $t$ under fixed $\varepsilon$ is bounded from above by
\begin{equation}\label{Eq6:Approximate_Query_Omega_1}
    \alpha t + \omega t \log (\omega t) + \omega t \log (1/\varepsilon).
\end{equation}
Due to the second quasi-linear term, this scaling is nearly-optimal rather than optimal in a precise sense.
However, the assumption $\omega \in \order{N^0}$ indicates that the second term cannot be dominant unless we consider inaccessible exponentially-large time scale $t \gtrsim e^{\alpha /\omega} T \in \order{e^{\poly{N}} T}$.
Therefore, the query complexity is optimal in time $t$ within practical usage of quantum computers.
On the other hand, the scaling in $\varepsilon$ is given by
\begin{equation}
    \omega t \frac{\log (1/\varepsilon)}{ \log \log \log (1/\varepsilon)},
\end{equation}
which is nearly-optimal.

When comparing the query complexity with those of other algorithms, we emphasize that our algorithm again achieves the additive scaling in the form of
\begin{equation}
    \poly{N}t + \omega t \times o(\log (1/\varepsilon)),
\end{equation}
where we neglect the second term in Eq. (\ref{Eq6:Approximate_Query_Omega_1}).
Although it has a product of time $t$ and the $o(\log (1/\varepsilon))$ term, it is sufficiently close to the theoretically-best scaling for time-independent systems given by Eq. (\ref{Eq2:query_qubitization}) due to the scale $\omega \in \order{N^0}$.
This also concludes that our algorithm saves the cost also for $\Omega (1)$-period dynamics compared to the Dyson-series algorithm yielding the productive scaling as Eq. (\ref{Eq6:Query_Dyson}).

\section{Discussion and Conclusion}\label{Sec:Discussion}

In this paper, we establish an efficient and accurate way to simulate multi-periodic time-dependent Hamiltonians, which include time-periodic and time-quasiperiodic systems.
Exploiting the block-encoding for each Fourier component as oracles, the algorithm can be executed with optimal/nearly-optimal query complexity in time $t$ and allowable error $\varepsilon$.
In addition, it achieves the query complexity with additive scaling sufficiently close to the best one for Hamiltonian simulation, and thereby deals with time-dependency much more efficiently than the truncated-Dyson-series algorithm.
While we assume the cutoff $m_\imax \in \order{1}$ for Fourier components $H_{\vec{m}}$, we expect the same results for multi-periodic Hamiltonians with exponentially-decaying Fourier components such that $\norm{H_{\vec{m}}} \leq e^{- \order{|\vec{m}|}}$ in a similar way to Ref. \cite{Mizuta2022optimal}.
Our result provides one significant step for understanding the complexity of time-dependent Hamiltonian simulation.
Furthermore, it will serve a promising application of quantum computers for condensed matter physics and quantum chemistry; for instance, it will be useful for exploration of nonequilibrium phenomena in quantum materials \cite{Chu2004-we-multi,Martin2017-multi,Crowley2019-multi,Zhao2019-multi,Else2020-multi} or preparing preferable quantum states by adiabatic dynamics \cite{Albash2018RevModPhys}.

We end up with leaving some future directions.
Our results play a role in extending a class of efficiently-simulatable nonequilibrium systems to multi-periodic time-dependent systems.
It is of great importance whether Hamiltonian simulation for other or all the time-dependent systems can reach the theoretically-best additive query complexity in time $t$ and allowable error $\varepsilon$.
Our success relies on the fact that their time-dependency is designated by $ \{ e^{i \vec{m} \cdot \vec{\omega} t} \}_{\vec{m}}$.
This results in a feasible extended Hilbert space equipped with Fourier indices $\{ \ket{\vec{l}} \}_{\vec{l}}$, which can be dealt with QET.
A possible direction for simulating other time-dependent Hamiltonians may be to consider those described by a finite set of basis functions like $ e^{i \vec{m} \cdot \vec{\omega} t}$.
Once one can find a proper ancilla degrees of freedom and an effective Hamiltonian for them, we expect that their Hamiltonian simulation can be accelerated by QET.

We also expect that our algorithm can be a clue to efficiently implementing multi-variable functions of matrices based on QET or QSVT.
In the present stage, dealing with multi-variable functions is difficult, except for cases where a set of matrices commutes with one another \cite{Salinas2021-approximant,Rossi_2022}.
Our algorithm can be viewed as the realization of the time-evolution $U(t)$, which nontrivially depends on the non-commutative variables $\{ H_{\vec{m}} \}_{\vec{m} \in M}$.
The key ingredient for this success is to embed multiple variables into a single variable in an extended Hilbert space, and to extract a desirable solution from it.
We hope that this strategy can be exploited for QET or QSVT toward some other multi-variable functions in broad fields.

\section*{Acknowledgment}

We thank K. Fujii for fruitful discussion.
K. M. is supported by RIKEN Special Postdoctoral Researcher Program.
\bibliography{bibliography.bib}

\appendix
\begin{center}
\bf{\large Appendix}
\end{center}

\section{Lieb-Robinson bound in Floquet-Hilbert space}\label{Sec:Lieb_Robinson}

Here, we prove the inequality Eq. (\ref{Eq4:Lieb_Robinson}) stated in Section \ref{Subsec:time_evol_repr}.
The proof is completely similar to that for time-periodic Hamiltonians, given by Ref. \cite{Mizuta2022optimal}.
The precise statement and its proof for multi-periodic time-dependent Hamiltonians are given as follows.

\begin{theorem}\label{Thm:Lieb_Robinson_bound}
\textbf{(Lieb-Robinson bound)}

 We impose all the assumptions on multi-periodic Hamiltonians $H(t)$ described in Section \ref{Subsec:Setup}.
 We define the parameter $\gamma$ by
\begin{equation}\label{EqA:def_gamma}
    \gamma = \sup_{\vec{x} \in [0,2\pi)^n} \left( \norm{\overline{H}(\vec{x}) - H_{\vec{0}}}\right) \leq \alpha.
\end{equation}
We consider two points $\vec{l},\vec{l}^\prime \in [L]^n$, and define their distance by that on the $n$-dimensional torus $[L]^n$ as
\begin{equation}
    d^L(\vec{l},\vec{l}^\prime) = \sqrt{\sum_{i=1}^n \left( \min \{ |l_i-l_i^\prime|, 2L - |l_i-l_i^\prime| \} \right)^2}.
\end{equation}
 Then, the transition amplitude is bounded from above by
 \begin{equation}\label{EqA:Lieb_Robinson}
     \norm{\braket{\vec{l}|e^{-i \ms{H}^L t}|\vec{l}^\prime}} \leq \left( \frac{e m_\imax \gamma t}{d^L(\vec{l},\vec{l}^\prime)} \right)^{d^L(\vec{l},\vec{l}^\prime|)/m_\imax},
 \end{equation}
in case $d(\vec{l},\vec{l}^\prime) \geq 2 m_\imax \gamma t$.
\end{theorem}

\textit{Proof.---} We split the effective Hamiltonian by $\ms{H}^L = \ms{H}_0 + \ms{H}_I$ with
\begin{eqnarray}
\ms{H}_0 &=& \sum_{\vec{l} \in [L]^n} \ket{\vec{l}}\bra{\vec{l}}_f \otimes (H_{\vec{0}} - \vec{l}\cdot \vec{\omega} I), \label{EqA:def_H_0}\\
\ms{H}_I &=& \sum_{\vec{m} \in M; \vec{m} \neq \vec{0}} \mr{Add}_{\vec{m}}^L \otimes H_{\vec{m}}. \label{EqA:def_H_int}
\end{eqnarray}
We employ the interaction picture based on the unitary transformation $\ms{H}_I(t) = e^{i \ms{H}_0 t } \ms{H}_I e^{-i \ms{H}_0 t}$.
The transition amplitude is evaluated by the Dyson-series expansion as 
\begin{eqnarray}
    && \norm{\braket{\vec{l}|e^{-i \ms{H}^L t}|\vec{l}^\prime}} \nonumber \\
    && \quad \leq \sum_{j=0}^\infty \int_0^t \dd t_j \hdots \int_0^{t_2} \dd t_1 \norm{\braket{\vec{l} |\ms{H}_I(t_j) \hdots \ms{H}_I(t_1)| \vec{l}^\prime}_f}. \nonumber \\
    && \label{EqA:transition_bound_1}
\end{eqnarray}
By inserting the identity $\sum_{\vec{l}_{j^\prime} \in [L]^n} \ket{\vec{l}_{j^\prime}}\bra{\vec{l}_{j^\prime}}_f \otimes I$, each integrand is equal to
\begin{eqnarray}
&& \braket{\vec{l} |\ms{H}_I(t_j) \hdots \ms{H}_I(t_1)| \vec{l}^\prime}_f \nonumber \\
&& \quad = \sum_{\vec{l}_1,\hdots,\vec{l}_{j-1}} \left( \prod_{j^\prime=0}^{j-1} \braket{\vec{l}_{j^\prime+1} |\ms{H}_I(t_{j^\prime})|\vec{l}_{j^\prime}}\right),
\end{eqnarray}
where we fix $\vec{l}_0 = \vec{l}^\prime$ and $\vec{l}_j = \vec{l}$.
Each term in the right hand side means transition amplitude from $\vec{l}^\prime$ to $\vec{l}$ with $j$-time jumps via the path $\vec{l}^\prime \to \vec{l}_1 \to \hdots \to \vec{l}_{j-1} \to \vec{l}$ under the Hamiltonian $\ms{H}_I(t)$.
From the definitions denoted by Eqs. (\ref{EqA:def_H_0}) and (\ref{EqA:def_H_int}), $\ms{H}_I(t)$ can induce the shift of $\vec{l}_{j^\prime}$ by $\vec{m} \in M$ on the torus $[L]^n$.
Since we assume $|\vec{m}| \leq m_\imax$ for every $\vec{m} \in M$, we obtain
\begin{equation}
    \braket{\vec{l} |\ms{H}_I(t_j) \hdots \ms{H}_I(t_1)| \vec{l}^\prime}_f = 0, \quad \text{if} \quad j m_\imax < d^L (\vec{l},\vec{l}^\prime),
\end{equation}
with the usage of the distance on the torus, $d^L$.
Equation (\ref{EqA:transition_bound_1}) results in 
\begin{eqnarray}
    && \norm{\braket{\vec{l}|e^{-i \ms{H}^L t}|\vec{l}^\prime}} \nonumber \\
    && \quad \leq \sum_{j= \lceil d^L(\vec{l},\vec{l}^\prime)/m_\imax \rceil}^\infty \int_0^t \dd t_j \hdots \int_0^{t_2} \dd t_1  \prod_{j^\prime=1}^j \norm{\ms{H}_I(t_j)} \nonumber \\
    && \quad \leq \sum_{j= \lceil d^L(\vec{l},\vec{l}^\prime)/m_\imax \rceil}^\infty \frac{(\norm{\ms{H}_I} t)^j}{j!}.
\end{eqnarray}

We next evaluate the operator norm of $\ms{H}_I$.
This is a kind of multi-dimensional circular matrix \cite{Gray2006-da}, according to the definition Eq. (\ref{EqA:def_H_int}).
We switch the basis from $\{ \ket{l}_f \}_{\vec{l} \in [L]^n}$ to $\{ \ket{x}_f \}_{\vec{x} \in [L]^n}$ by Fourier series as
\begin{equation}
    \ket{\vec{x}}_f = \sum_{\vec{l} \in [L]^n} e^{i 2 \pi (\vec{l} \cdot \vec{x}/ 2L)} \ket{l}_f.
\end{equation}
Then, the Hamiltonian $\ms{H}_I$ is block-diagonalized in this basis as
\begin{eqnarray}
    \ms{H}_I &=& \sum_{\vec{x} \in [L]^n} \ket{\vec{x}}\bra{\vec{x}}_f \otimes \left( \sum_{\vec{m} \in M; \vec{m} \neq \vec{0}} H_{\vec{m}} e^{-i 2\pi (\vec{m} \cdot \vec{x} / 2L)} \right) \nonumber \\
    &=& \sum_{\vec{x} \in [L]^n} \ket{\vec{x}}\bra{\vec{x}}_f \otimes \left( \overline{H}\left( \frac{\pi \vec{x}}{L}\right) - H_{\vec{0}}\right).
\end{eqnarray}
We obtain $\norm{\ms{H}_I} \leq \gamma$ from its definition Eq. (\ref{EqA:def_gamma}).
Under the assumption $d^L(\vec{l},\vec{l}^\prime) \geq 2 m_\imax \gamma t$ and the notation $j_\imin = \lceil d^L (\vec{l}, \vec{l}^\prime) / m_\imax \rceil$, this relation leads to
\begin{eqnarray}
    \norm{\braket{\vec{l}|e^{-i \ms{H}^L t}|\vec{l}^\prime}} &\leq& \sum_{j= j_\imin}^\infty \frac{( \gamma t)^j}{j!} \nonumber \\
    &\leq& \frac{(\gamma t)^{j_\imin}}{j_\imin !} \sum_{j=j_\imin}^\infty \left( \frac{1}{2} \right)^{j-j_\imin} \nonumber \\
    &\leq& 2 \frac{(\gamma t)^{j_\imin}}{j_\imin !}.
\end{eqnarray}
Using the inequality $2x^j / j! \leq (ex / j)^j$ for $x>0$ and $j \in \bbN$, originating from the Stiring formula, we arrive at the inequality Eq. (\ref{EqA:Lieb_Robinson}), which completes the proof. $\qquad \square$

We briefly explain the origin of this bound.
The effective Hamiltonian $\ms{H}^L$ can be viewed as the one describing quantum walk on an $n$-dimensional lattice.
In this picture, each Fourier index $\vec{l} \in [L]^n$ and each Fourier component $H_{\vec{m}}$ respectively play roles of a lattice site and hopping by a vector $\vec{m}$.
The assumption $|\vec{m}| \leq m_\imax$ implies that the hopping is finite-ranged.
As a result, the inequality Eq. (\ref{EqA:Lieb_Robinson}) corresponds to the Lieb-Robinson bound on the transition amplitude of a $n$-dimensional quantum walk with finite-ranged hopping \cite{Lieb1972-uo,Gong2022-bound}, and this is why we refer to Theorem \ref{Thm:Lieb_Robinson_bound} as the Lieb-Robinson bound.

Similarly, when we assume $\norm{H_{\vec{m}}} \leq e^{-\order{\vec{m}}}$ for every $\vec{m}$ instead of $H_{\vec{m}}=0$ for $|\vec{m}| > m_\imax$, we expect the Lieb-Robinson bound corresponding to an $n$-dimensional quantum walk with exponentially-decaying quantum walk as
\begin{equation}
    \norm{\braket{\vec{l}|e^{-i \ms{H}^L t}|\vec{l}^\prime}} \leq \mr{Const.} \times e^{ \tilde{\alpha} t - d^L(\vec{l},\vec{l^\prime})/\xi}
\end{equation}
with some constants $\tilde{\alpha} \in \poly{N}$ and $\xi \in \order{1}$ \cite{Robinson1976-wd,Nachtergaele2006-ok}.
As discussed in Ref. \cite{Mizuta2022optimal} for time-periodic Hamiltonians, this relation leads to a proper cutoff $L \in \order{\tilde{\alpha} t + \log (1/\varepsilon)}$ for the Floquet-Hilbert space to accurately reproduce the dynamics within the error $\order{\varepsilon}$.
Therefore, as a consequence of universality of the Lieb-Robinson bound, we expect that our protocol achieves the additive query complexity in the form of $\order{\alpha t + \log (1/\varepsilon)}$ [$\omega t \in \order{1}$] or $\order{\alpha t + \omega t \log (\omega t/\varepsilon)}$ [$\omega t \in \Omega (1)$] also for time-dependent Hamiltonians satisfying $\norm{H_{\vec{m}}} \leq e^{-\order{|\vec{m}|}}$.

\end{document}